\documentclass{iopart}
\usepackage{amssymb,amsfonts,amstext}
\usepackage{hyperref,epsfig}
\usepackage[svgnames]{xcolor}
\usepackage{pgf,tikz}

\newcommand{\klmt}{\mbox{K\hspace{-7.6pt}KLM\hspace{-9.35pt}MT}\ }

\newcommand\beq{\begin{equation}}
\newcommand\eeq{\end{equation}}
\newcommand\bea{\begin{eqnarray}}
\newcommand\eea{\end{eqnarray}}

\newcommand{\OL}[1]{ \hspace{1pt}\overline{\hspace{-.5pt}#1
    \hspace{-1.5pt}}\hspace{1.5pt} }

\renewcommand{\vec}[1]{\mathbf{#1}}

\begin{document}

\review{Seeking String Theory in the Cosmos}
\author{Edmund~J.~Copeland$^{a}$, Levon~Pogosian$^{b}$, and Tanmay~Vachaspati$^{c}$}
%\address{
%  Department of Physics,
%  Simon Fraser University\\
%  8888 University Drive,
%  Burnaby, BC Canada
%  V5A 1S6
%}

\address{$^a$School of Physics and Astronomy, University of Nottingham, University Park, Nottingham NG7 2RD, UK\\
$^b$ Department of Physics, Simon Fraser University, Burnaby, BC, V5A 1S6, Canada \\
$^c$ Department of Physics, Arizona State University, Tempe, AZ 85287, USA
}
\ead{ed.copeland@nottingham.ac.uk, levon@sfu.edu, tvachasp@asu.edu}

\date{\today}

\begin{abstract}
We review the existence, formation and properties of cosmic strings in 
string theory, the wide variety of observational techniques that
are being employed to detect them, and the constraints that current
observations impose on string theory models. 
\end{abstract}

\submitto{\CQG \hspace{-0.2em}(special cluster issue)}
\maketitle

\section{Introduction}

It is clear that the standard model of particle physics is incomplete.
Observations and experiments provide evidence for dark forms of matter
and neutrino masses. Theoretically it is hard to understand the hierarchies
among standard model parameters, and the cosmic matter-antimatter asymmetry.
The standard model makes no attempt to describe a quantum theory of
gravity. Grand Unified models, with and without supersymmetry, try to
resolve the particle physics difficulties in a coherent framework, but
string theory is the only candidate for an all encompassing framework
that includes gravity.

The traditional approach to test models of the fundamental interactions
is to do collider experiments at higher energies and higher luminosities, 
thereby probing more massive states and weaker interactions. Over the last
several decades, a complementary approach to the study of fundamental
interactions has emerged. The approach recognizes that high energy
interactions were rampant in the early universe and, depending on the
nature of those interactions, relics from those epochs could have
survived until the present. Thus, to study the fundamental interactions
one could also look out in the universe and search for relics.

Cosmic strings are relics from the early universe and exist in a wide
variety of Grand Unified models of particle physics. In this context,
they exist for the same reasons that vortex solutions exist in superfluids
and superconductors. It was somewhat of a surprise to discover that cosmic 
strings can also exist in string theory. Early work on string theory cosmic
strings showed them to be heavy and unstable,
but as string theory models got more sophisticated, light and
stable cosmic strings were discovered. Hence, the search for
cosmic strings can potentially lead to the validation of string theory.
However, there are many steps that will be required of a rigorous
validation of string theory: not only do cosmic strings have to
be discovered but they also have to show signatures unique to
string theory.

Remarkable advances in observational cosmology over the last few
decades give a glimmer of hope that cosmic strings may be discovered,
or even if they are not discovered, may be constrained sufficiently so
as to give useful information for string theory model building. These
observational advances include high precision measurements of the
cosmic microwave background, enormous surveys that map out cosmic
large-scale structure, cosmic ray detectors with collection area
comparable to (small) countries, satellite observations of high
energy gamma ray photons, neutrino detectors that dig miles into 
the pure Antarctic ice, searches for and discovery of many more 
millisecond pulsars, and sophisticated gravitational wave detectors.

Our aim in this review is to describe the basic conditions necessary
for the existence of cosmic strings in string theory models and to
outline the current observational constraints on cosmic string
parameters. In some cases, there is uncertainty about the constraints
because properties of the network of cosmic strings  or their interactions
with matter are not completely understood. We will try and clarify
where such uncertainty exists.

In Sec.~\ref{sec:existence} we describe cosmic string solutions
in string theory. This is a rich subject since there are two
different strings (F- and D-strings) and they can form bound
states and junctions. In certain string theory motivated field
theory models, those with bifundamental fields, similar strings
can arise and these are also discussed. In Sec.~\ref{sec:observation}
we describe the many different cosmological tools that are
available to look for cosmic strings. Here we also discuss current
constraints. We summarize our discussion in Sec.~\ref{sec:summary}.

\section{Cosmic strings in string theory}
\label{sec:existence}

\subsection{F- and D-strings}
\label{subsec:FD}

The early development of superstrings actually coincided with that of cosmic strings,  and lead Witten to ask whether they could be related \cite{Witten:1985fp}. The usual expectation of superstrings was that they were of order the Planck length, so microscopic in size. Was there any possibility that these strings created in the early universe could have been stretched by the cosmological expansion to reach macroscopic sizes, which would have then been seen as cosmic strings?  Witten's answer was in the negative. First of all fundamental
strings were expected to have tensions $\mu$ close to the Planck scale, for example in
perturbative heterotic string theory, $G \mu = \alpha_{\rm GUT} / 16\pi \geq 10^{-3}$, whereas the isotropy of the
cosmic microwave background constrained any string of
cosmic size to satisfy $G\mu \leq 10^{-5}$, inconsistent with them being fundamental strings. Moreover given that fundamental strings would have been produced in the early
universe, they would have been diluted by a period of  cosmological inflation, so rather than being stretched to macroscopic size with the expansion of the
universe, they would have been diluted away (as were the monopoles) leaving a universe effectively empty of Planck scale cosmic fundamental strings.  With regard to their formation in the first place, Witten showed that macroscopic
Type I strings break up on a stringy time scale into short open strings,
and so would never form.  It was known that macroscopic heterotic strings appear as
boundaries of axion domain walls, whose tension would force the strings to
collapse rather than grow to cosmic  scales \cite{Vilenkin:1982ks} . For Type II strings it is the Neveu-Schwarz 5-brane instantons, (in combination with
supersymmetry breaking to lift the zero modes) that produce an axion
potential,  leading to domain walls and the resultant collapse of the strings
\cite{Becker:1995kb}. 

Since the second string revolution, post 1995, it has become clear that string theory has a much richer spectrum of objects than simple fundamental strings. Many new one-dimensional objects are now known: in addition to the fundamental
F-strings, there are D-strings (D- for Dirichlet), as well as  higher dimensional D-, NS- M-branes that are partially wrapped on compact cycles so that only one
noncompact dimension remains.  The possibility that the extra compact dimensions could be large \cite{ArkaniHamed:1998rs,ArkaniHamed:1998nn,Antoniadis:1998ig}  and/or warped along some of the internal dimensions \cite{Randall:1999ee,Randall:1999vf}, has allowed for the existence of much lower tensions for these strings which could be out there in the universe waiting to be detected.  For large extra dimensions, using Kaluza-Klein reduction we can relate the fundamental 10 dimensional Newton's constant  $G_{10}$ and the effective 4 dimensional one we perceive $G_4$, through $G_4=G_{10}/V_6$ where $V_6$ is the volume of the six internal compact dimensions of space time. It implies that the four-dimensional Planck length is derived from the fundamental ten dimensional Planck length, hence when $V_6 \gg l^6_s$ (where $l_s$ is the String length), which corresponds to the large extra dimension scenarios, the observed four dimensional Planck length is much smaller than the fundamental higher dimensional Planck length. The fundamental string tension is then much smaller than the observed four dimensional Planck scale, $\mu_{\rm Fun} = {1\over 2\pi l^2_s} \ll {1\over l^2_{\rm 4d-Pl}} $, which of course implies that we can obtain a low tension in large extra dimensional scenarios. Another approach involves introducing a warping factor, where the internal space impacts directly on the four dimensional metric and is usually given by a contribution to the line element of the form $e^{2A(y)} g_{\mu \nu} dx^\mu dx^\nu$ where $y$ is the internal dimension. Now in some compactifications there exist throat regions where the warping can be very strong, $e^{2A_0} \ll 1$. These scenarios tend to arise from models with fluxes on the internal manifold \cite{Giddings:2001yu}. If a fundamental string falls to the bottom of the throat, it will have an effective tension which is much less than the fundamental string tension $\mu_{\rm Fun}$ as perceived in the bulk, i.e. $\mu_{\rm eff} = e^{2A_0} \mu_{\rm Fun} \ll \mu_{\rm Fun}$, but its value will depend on the internal space coordinate. It is this gravitational potential of the warped throat which ensures that strings which fall into the throat remain there.  However, there are constraints. For any given geometry the form of the brane-antibrane potential is known,
and there exists a relation between the observed magnitude of
density fluctuations $\delta_H$ and the parameters of the model.  For the
models of \cite{Jones:2002cv,Jones:2003da,Sarangi:2002yt}, which are based on unwarped compactifications
with the moduli fixed by hand, the authors find a range
$10^{-11} \leq G\mu \leq 10^{-6}$, with a narrower range around $10^{-7}$ for their favoured models based on
branes at small angles. The key point is that at least based on allowed tensions, the existence of cosmic superstrings can be perfectly consistent with the observed anisotropy of the cosmic microwave background radiation. 

One of the primary routes for the formation of D-strings is upon the exit from inflation in D-brane inflation scenarios involving collisions of D3- branes and anti-D3-branes  \cite{Burgess:2001fx}. In \cite{Jones:2002cv,Jones:2003da,Sarangi:2002yt} the authors argued that D-brane-antibrane inflation 
\cite{Dvali:1998pa,Burgess:2001fx,Alexander:2001ks,Dvali:2001fw},  lead to the production of lower-dimensional D-branes that are one-dimensional in the noncompact directions.  In \cite{Jones:2002cv,Sarangi:2002yt} the authors  made the important observation that zero-dimensional defects (monopoles) and two-dimensional defects (domain walls) are not produced; a fortunate result as either of these would have led to major cosmological problems. This type of inflation can be considered as the string realisation of hybrid inflation which in this case produces D-strings at the end of inflation. The
D1-branes are topological defects in the tachyon field mediating
D3-$\OL{\rm D3}$ annihilation \cite{Sen:1999mg}, and 
their production can be described by the standard Kibble mechanism \cite{Kibble:1976sj}. The fundamental strings do not have a classical description in terms of the same variables, but there are dualities present, in this case S-duality that relates the production of D-strings in the collision of
a D3-brane with an anti-D3-brane to the production of fundamental strings, implying that both D-strings and F-strings are expected to be produced at the end of brane inflation \cite{Copeland:2003bj,Dvali:2002fi,Dvali:2003zj}. Both D and F- type cosmic strings are produced because of the way two symmetries are broken during inflation.  By the time inflation ends the worldvolume gauge symmetries of both the D3 and the anti-D3-brane have been broken, in particular two $U(1)$ symmetries have been broken.  Recall that in this scenario, all the strings created at the end of inflation are at the bottom of the inflationary throat, and will remain there because they are in a deep potential well.  Although only two types of string are produced, we end up with a much richer spectrum of string types (unlike conventional abelian cosmic strings), because when a D string meets an F-string it can bind together rather than intercommute. In other words they can merge to form
a bound state known as a (1, 1)-string. Further binding leads to the formation of higher bound states, generally known as (p, q)-strings, which are composed of p F-strings and q D-strings where p and q are relatively prime integers. They integers need to be coprime in order to ensure the bound states are stable and are now interpreted  as bound states of $p$ F1-branes and $q$ D1-branes \cite{Polchinski:1995mt,Witten:1995im}.
Their tension in the ten-dimensional type IIb theory is \cite{Schwarz:1995dk}
$\bar \mu_{p,q} = \frac{1}{2\pi l^2_s} \sqrt{ p^2 + \frac{q^2}{g_{\rm s}^2}}$ where $g_{\rm s}$ is the perturbative string coupling.  This result is valid for (p,q)-bound states in a flat ten dimensional spacetime.

In general the D-brane inflation models involving the collision of D3-anti D3 branes leads to the production of D, F, and combined (p,q) strings, which if stable enough could still be present in the universe today as cosmic strings. It is possible that multiple throats exist in the compact space which then allows for the possibility of decoupling the high energy inflation scale (associated with motion in one throat) and the lower scales relevant for particle phenomenology (including the formation of strings) in one of the other deeper throats. The key idea is that reheating at the end of inflation could propagate over to the deeper throat, leading to an increase in temperature in that throat above the deconfinement temperature. The universe then cools back to its confined state again, with the inevitable production of cosmic superstrings at this lower energy scale \cite{Horowitz:2007fe}. In general  for cosmic superstrings to form in models of D-brane inflation, inflation must end with the annihilation of some of the space-filling branes. Although, strings are not an inevitable consequence of brane inflation,
%They do not have to form if there is no annihilation of the space filling branes at the end of inflation. An example of this is the condensation of an open string tachyon, and in many models involving inflation based on closed string moduli. However, 
given the expectation that the  particle phenomenology associated with superstrings is expected to be rich, we may well expect there to be standard cosmic strings, with tensions well below the string scale emerging out of supersymmetric grand unified theories for example \cite{Jeannerot:2003qv}. 

We finish this section with a brief word about the instabilities associated with F- and D-strings. F-strings are unstable to both fragmentation in open string theories and confinement by axion domain walls with the resulting wall tension causing the string loops to collapse. There are more {\em potential} instabilities including the two Witten pointed out which we now think of in terms of the breakage of strings on space-filing branes and the confinement of axion domain walls, as well as two new ones, namely an effect similar to baryon decay and tachyon condensation. They are technical calculations and we will not go into any details in this review, the interested reader is recommended to read \cite{Myers:1900zz,Polchinski:2004ia,Polchinski:2004hb,Copeland:2003bj}. The bottom line is that there are clear modes of decay for F and D strings in all the models, and for them to remain viable candidates for cosmology we have to be lucky with the conditions so as to suppress the natural decay routes. In the next section we will turn our attention to two such models. 
 
\subsection{Models leading to the formation of cosmic superstrings}
\label{str-models}

\subsubsection{The \klmt model}
\label{klmt}
The \klmt  model  \cite{Kachru:2003sx} was initially introduced as a way of realising inflation in string theory in a framework where all the moduli are stabilised \cite{Kachru:2003aw}. The realisation that cosmic superstrings formed at the end of inflation is an extra exciting feature of the model, which is based on IIB string theory on a Calabi-Yau manifold. The  Calabi-Yau is orientifolded by a ${\bf Z}_2$ symmetry with isolated fixed points, which become O3-planes. The spacetime
metric is warped, with the inflaton being the separation between a D3-brane and an anti-D3-brane,
whose annihilation leads to reheating.  The annihilation occurs in a region
(throat) of large gravitational redshift,  although the majority of the bulk of the  Calabi-Yau does not experience such dramatic warping. The redshift in the throat plays a key role: both the inflationary scale and the scale of string tension, as measured by a ten-dimensional inertial observer, are governed by string physics and are close to the four-dimensional Planck scale, but the corresponding energy scales as measured by a four-dimensional physicist are suppressed by the large warping factor. As discussed in \cite{Jones:2002cv,Jones:2003da,Sarangi:2002yt} we expect that  only one-dimensional objects in the non compact dimensions will be produced in any significant numbers that will lie entirely within the region of reheating.  The obvious candidates are then the F1-brane (fundamental IIB string) and D1-brane, localized in the throat.  As mentioned earlier, the D1-branes can be regarded as topological defects in the tachyon field
that describes D3-$\OL{\mbox{D3}}$ annihilation \cite{Sen:1999mg,Witten:1998cd,Horava:1998jy}, and
so these will be produced by the Kibble mechanism \cite{Kibble:1976sj}.  The F1-branes do not
have a classical description in these same variables, but in an $S$-dual
description they are topological defects and so must be produced in the
same way.  Of course, only one of the $S$-dual descriptions can be
quantitatively valid, and if the string coupling is of order one then
neither is.  However, given that the Kibble argument depends only on causality it is probably valid for both kinds of string in all regimes. 

We have argued that the \klmt model leads to the formation of strings, but we need to know about their stability. This depends in what branes remain in the theory after inflation for the $p,q$ strings to break on (and hence decay). Now because in the \klmt model there need to be branes to host the standard model fields, and there need to be extra anti D3-branes located in the throat to ensure the moduli are stabilised, it of course implies there are extra branes present after inflation. We won't go into details here of the consequence of these branes and the conditions that need to be satisfied for the strings formed to be meta-stable (i.e. have a lifetime at least the age of the Universe), but we direct the interested reader to  \cite{Myers:1900zz,Polchinski:2004ia,Polchinski:2004hb,Copeland:2003bj} for the technical details. Basically, we just need to know that there are conditions under which the strings formed can be stable against decay and therefore remain cosmologically interesting for us.

\subsubsection{Large dimension models}
\label{large-dim}

In the \klmt model the string scale is lowered by a large warp factor.  As we have discussed earlier, it can
also be lowered in the context of large compact dimensions without such a large warping as first shown in  \cite{Jones:2002cv,Jones:2003da,Sarangi:2002yt}.
Although in these models there are not yet examples with all moduli
stabilized, it is still possible to investigate the stability of potential cosmic strings by first of all 
fixing the moduli by hand. When this is done, a rich spectra of strings is found depending on 
the compactification scenario. For example for the case of wrapped branes in type II compactification, in \cite{Jones:2002cv,Jones:2003da,Sarangi:2002yt}  the authors found a range of strings in the range $10^{-12} \leq G\mu \leq 10^{-6}$, assuming inflation is responsible for the generation of the CMB anisotropies.  

\subsection{Intercommuting properties}
\label{reconnect}
In this section we begin to address a vital question concerning cosmic superstrings. How can we differentiate a network of them from the more traditional field theory based cosmic strings? We have seen in earlier sections that it is possible to form cosmic superstrings from string theory and that they can be long lived (cosmologically), survive  a period of inflation and have a macroscopic length. They bring with them two particular features that may help us distinguish the two types. They are a reduced probability of intercommuting \cite{Jackson:2004zg, Jackson:2007hn} and the formation of junctions in the (p,q) type networks \cite{Sen:1999mg}. 
We will concentrate on these two features in this section.

Field theory simulations of cosmic strings indicate that the probability of intercommuting is essentially unity, $P=1$, with only ultra-relativistic strings being able to pass through each other without reconnecting, a result true of both global \cite{Shellard:1987bv} and local \cite{Matzner:1989} cosmic strings. The case for superstrings is different however, as pointed out in great detail in \cite{Jackson:2004zg, Jackson:2007hn}. Polchinski \cite{Polchinski:1988cn} had earlier shown that for fundamental strings, the reconnection probability depended primarily on the string coupling constant $g_s$ and is of order $g^2_s$ allowing it  to be much less than one. The work in  \cite{Jackson:2004zg} extended this approach analysing the collisions of fundamental and Dirichlet strings, as well as  their (p, q) bound states and also between all possible pairs of strings. The strength of the interaction between the colliding strings  depend on the details of compactification; the relative velocity of the strings, their intersection angle and crucially on $g_s$. Basically the probability of reconnection is generally below unity implying that F- and D-strings can in principle be distinguished from gauge theory strings. In fact in some cases $P\sim 10^{-3}$ which would have a large effect on the behavior of string
networks. For strings of different types the reconnection probability depends strongly on the details of the compactification. It can be large or essentially zero.  An important consequence of a reduction in intercommutation rates is that  the density of long strings has to increase because loop formation becomes less efficient as a mechanism for energy loss. 
%\subsubsection{Bound states} 
%\label{bound-states}

When strings of two different types cross they cannot intercommute in the
same way as usual.  Rather they can produce a pair of trilinear vertices 
connected by a segment of string. For example, the crossing 
of a $(p,q)$ string and a $(p',q')$ string can produce a $(p+p',q+q')$ 
string or a $(p-p',q-q')$ string. Only for 
$(p',q') = \pm (p,q)$ is the usual intercommutation possible. Now, we saw
earlier that the D3 anti-D3 brane inflation model can lead to 
the formation of a network of (p,q)-strings. In general the string tension depends on the square root of a function 
of p and q squared. Over the past few years an alternative approach 
has emerged to determine the properties of a network of (p,q) strings, which should be a good approximation at least for the lowest lying tension states. That has been the development of field theory analogues, either combinations of interacting abelian models or through non-abelian models . In both cases the networks that form admit trilinear vertices, hence junctions where the usual intercommutation properties of the strings no longer applies. As mentioned earlier, evidence appears to indicate that at least for these lowest lying string tensions, the networks that form reach a scaling regime . Similar results are obtained in analytical approaches which have been developed (for more details and detailed references see  \cite{Copeland:2009ga}). 

A key aspect of string theory is the existence of string dualities which relate different string models, hence can relate different types of string that are formed. This means that the spectrium of bound state strings can be much richer than just the (p,q)-strings discussed so far. The (p,q)-strings form a mulitiplet under a discrete SL(2,Z) symmetry of the ten-dimensional type IIb theory\cite{Schwarz:1995dk}, which means that an F-string or (1,0)-string can be mapped to a general (p,q)-string by the application of an appropriate SL(2,Z) transformation, leading to the formation of many types of string, not just the basic F-string. However, it does not mean they would all form, that would depend on the detailed dynamics of the particular model being considered, but in principle they could form leading to a rich structure of low tension bound cosmic superstrings. 

\subsection{Cosmic superstring scaling solutions}

We turn our attention to the nature of the scaling solutions found in a network of cosmic superstrings. Not surprisingly, 
the evolution of a network that contains bound state 
strings is different from that of a network without bound states. 
The additional binding of the strings allows for more complicated 
configurations and leads to new energy loss mechanisms to the network. Several approaches have been developed to model the evolution of cosmic string networks, and an interesting recent attempt to extend them to cosmic superstring networks -- which contain different types of string -- is due to Tye, Wasserman and Wyman \cite{TWW}. Their model, based on the ``velocity dependent one-scale'' model of Martins and Shellard \cite{VOS,VOSk}, describes evolution of a multiple tension string network (MTSN) under the assumption that all types of strings have the same correlation length and root-mean-square velocity. By studying the evolution of the number density of strings, they find that scaling is achieved when the energy associated to the formation of junctions is assumed to be radiated away. This model has been extended in \cite{NAVOS}, where the authors assigned a different correlation length and velocity to each string type, and enforced energy conservation at each junction. Scaling is again achieved (with different number densities), but not as generically as in \cite{TWW}. 
The conclusion, based on these analytic and numerical approaches is that a network of bound strings will reach a scaling 
solution with the lowest lying states, the F-,D- and (1,1)-strings 
having the higher number densities compared to the higher tension strings. 
This is because the unbinding of higher tension strings is favoured 
kinematically compared with the binding processes. The kinematics of 
strings that can form junctions is a fascinating area of research which has 
only recently started receiving attention 
\cite{Copeland:2006eh,Copeland:2006if,Copeland:2007nv,Davis:2008kg}.  
A number of important features have emerged through studies of the 
collisions of Nambu-Goto strings with junctions at which three strings meet.  
One is that the exchange to form junctions cannot occur if the strings 
meet with very large relative velocity  
\cite{Copeland:2006eh,Copeland:2006if,Copeland:2007nv,Firouzjahi:2009nt}. 
For the case of non-abelian strings rather than passing through one 
another they become stuck in an X configuration \cite{Copeland:2006if}, 
in each case the constraint depends on the angle at which the strings meet, 
on their relative velocity, and on the ratios of the string tensions. 
Under the assumption that, in a network, the incoming waves at a 
junction are independently randomly distributed, it is possible to 
determine the r.m.s.\ velocities of strings and calculate the average 
speed at which a junction moves along each of the three strings from 
which it is formed \cite{Copeland:2006if}. The results are consistent with 
what we have mentioned above, namely that junction dynamics may be such 
as to preferentially remove the heavy strings from the network leaving 
a network of predominantly light strings.  In \cite{Copeland:2007nv} 
the authors modified the Nambu-Goto equations to include the formation 
of three-string junctions between $(p,q)$-cosmic superstrings, which 
required suitable modifications to take account of the additional 
requirements of flux conservation. Investigating the collisions between 
such strings they showed that kinematic constraints analogous to those 
found previously for collisions of Nambu-Goto strings apply here too. 
Extending their analysis to the \klmt motivated model of the formation 
of junctions for strings in a warped space, specifically with a 
Klebanov-Strassler throat, they showed that similar constraints still 
apply with changes to the parameters taking account of the warping and 
the background flux. 
%%%%%%%%%%%%%%%%%Title move
%In a complementary approach, a number of authors have studied the kinematics of cosmic string collisions \cite{CopKibSteer1,CopKibSteer2}. When two Nambu-Goto (NG) strings (of generally different tensions) collide, rather than intercommuting in the standard way, they can form two junctions and a linking string of a third tension. Kinematically this can only occur if the relative orientation, velocity and string tensions lie in certain ranges. In \cite{CopFirKibSteer}, the authors extended their earlier studies to $(p,q)$-cosmic superstrings by modifying the NG equations to take into account the additional requirements of flux conservation. Once again the kinematic conditions required for the formation of Y-junctions were established, with results very similar to the ones obtained for NG strings.  
These kinematic constraints have been checked quite extensively with dynamical field theory simulations of strings collisions, and the agreement is (generally) good \cite{Sakellariadou:2008ay,Salmi,Urrestilla:2007yw,Bevis:2008hg,Bevis:2009az}.  Recently, in \cite{Avgoustidis:2009ke} these constraints have been incorporated into the generalised MTSN velocity one-scale model of \cite{NAVOS}, leading to new conditions required for scaling and thereby providing the most complete model of cosmic superstring evolution to date. 

In~\cite{Pourtsidou:2010gu}, the model of \cite{Avgoustidis:2009ke} was used to study the evolution of a cosmic superstring network for different values of the string coupling $g_s$ and different charges $(p,q)$ on the strings. It was found that in all cases the three lightest strings, i.~e. the $(1,0)$, $(0,1)$ and $(1,1)$ strings, dominate the string {\it number} density. When the string coupling is large, $g_s \sim {\cal O}(1)$, most of the network {\it energy} density is in the lightest $(1,0)$ and $(0,1)$ strings (respectively F and D strings), whose tensions are approximately equal. At smaller values of $g_s \sim {\cal O}(10^{-2})$, the $(1,0)$ string becomes much lighter than both the  $(0,1)$ and $(1,1)$ strings, and dominates the string {\it number} density. However, because of their much larger tension, the {\it energy} density of the network at small couplings can be dominated by the rarer $(0,1)$ and $(1,1)$ strings. 
The existence of these two distinct limiting scaling behaviours at large or small values of $g_s$ is quite generic, although the specific details are somewhat dependent on the model-dependent value of the effective volume of the compactified dimensions. In either of the two limiting regimes, the {\it energy} density of the multi-tension network is effectively dominated by strings of one tension.

A possible problem for cosmic superstrings pointed out in \cite{Avgoustidis:2005vm} exists even in scenarios where monopoles and 
domain walls (the usual problem defects in cosmology) do not form. It 
involves the possibility that in winding around  compact extra dimension 
in a way that forbids the loop to vanish the strings themselves can form 
new stable remnants they term as {\it cycloops} and if they do form in 
a network the strings responsible for them would have to have incredibly  
small tensions of order $G\mu < 10^{-18}$, which would be impossible to 
detect with current or planned experiments. There also remains the 
possibility of stable loops of superconducting string forming 
{\it vortons} \cite{Davis:1989nj} which are stabilised by their angular 
momentum and can not be radiated away classically. If formed these would 
be disastrous cosmologically for $G\mu > 10^{-20}$ 
\cite{Martins:1998gb,Martins:1998th}. However these conclusions rely 
on the existence of superconducting currents that interact
with standard model gauge fields. This is not very likely, as we 
already have established that for cosmic superstrings to be stable 
against breakage on space filling branes, the condition seems to require 
that cosmic superstrings have only indirect, e.g., gravitational interactions 
with the Standard Model. Of course it is possible to turn these problems 
on their head and tune the parameters so that the remnant loops actually 
play the role of dark matter.

\subsection{Bifundamental strings}
\label{subsec:bifunda}

Several string theory models at low energies lead to ``bi-fundamental
matter'', that is fields that transform in the fundamental representations
of several different non-Abelian groups \cite{Berkooz:1996km,Katz:1997eq}.
To understand the prevalence of bi-fundamental matter in string theory, 
note that an open string has two ends and these
can be stuck to two separate stacks of branes. The strings then transform
under a group corresponding to one stack and also corresponding to
the other stack. An effective field theory description is in terms
of a field, call it $B$, that transforms in the fundamental representation
of two separate groups, call them $G_1$ and $G_2$.

\begin{figure}
\begin{center}
  \includegraphics[width=1.0in,angle=90]{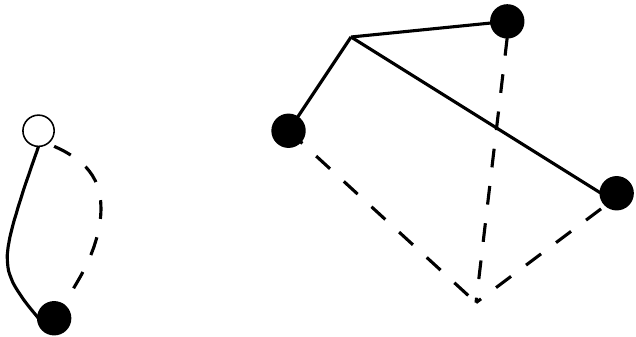}
\end{center}
\caption{Hadrons of bi-fundamentals can consist of a number of
bi-fundamental particles, each confined by two types of flux tubes
(solid and dashed).
}
\label{baryonmeson}
\end{figure}

\begin{figure}
\begin{center}
  \includegraphics[width=1.50in,angle=90]{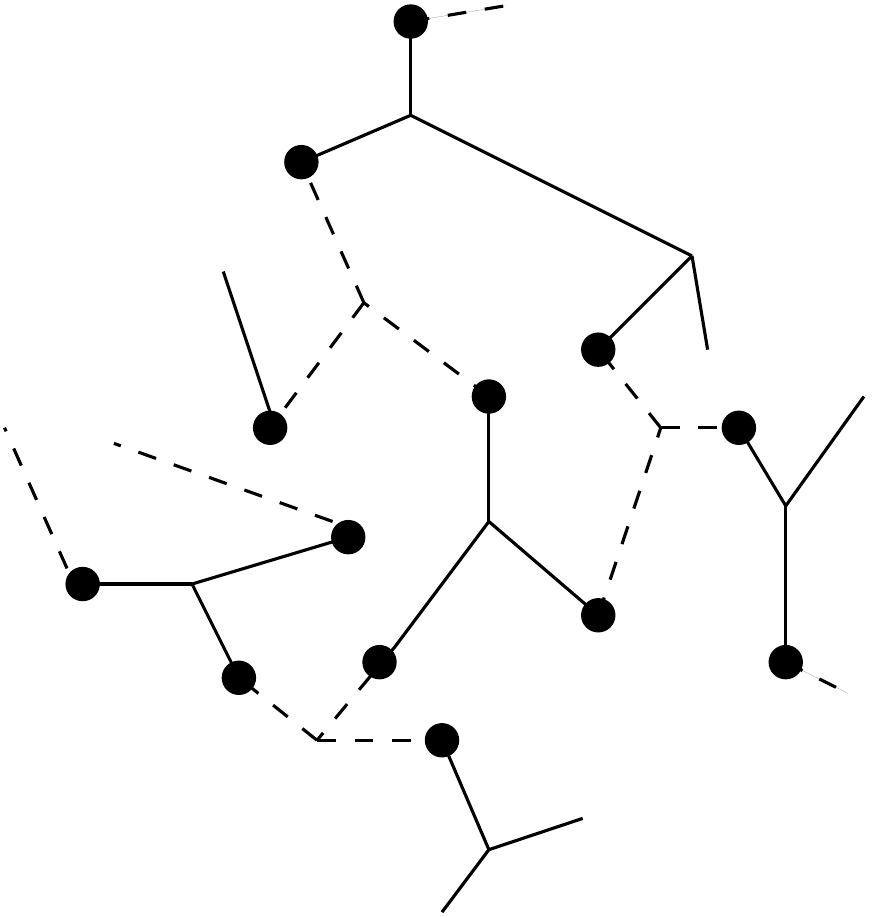}
\end{center}
\caption{The web of bi-fundamentals. Two types of flux tubes
connect each particle, and the network percolates, forming an
infinite web. The transition to a hadronic phase is frustrated.
}
\label{net}
\end{figure}

An interesting situation arises if the two symmetry groups are non-Abelian
and confining \cite{Vachaspati:2008wi}. In that case, particles of $B$
have to form singlets (hadrons)
of $G_1$ as well as $G_2$. This will happen by the formation of electric
flux tubes that confine the particles. Examples of some such hadrons are
shown in Fig.~\ref{baryonmeson} when $G_1$ and $G_2$ are both $SU(3)$.
However, experience with the formation of cosmic strings in the Abelian-Higgs
model suggests that the deconfinement to confinement transition
cannot lead to hadronization. Even though energy considerations
imply hadronization, it is entropically much more favorable for
the bi-fundamentals to form an infinite web of strings as shown
in Fig.~\ref{net}.

An implicit assumption in the above discussion is that the confining
electric strings cannot fragment. Fragmentation is possible if there
exists a light particle that transforms in the fundamental representation
of either $G_1$ or of $G_2$ and is a singlet under the other group.
This is precisely the situation in the standard model of electroweak
interactions. There we have left-handed quarks that are bi-fundamentals
of $G_1=SU(3)_c$ and $G_2=SU(2)_L$, but the model also contains
singlets of $G_2$, namely the right-handed quarks, as well as
the leptons that are singlets under $G_1$. Thus the confining
strings are unstable to fragmentation. Also, in the standard model,
$SU(2)_L$ gets spontaneously broken and the electroweak sector is 
never in a confining phase.

This shows that a cosmic network of electric strings can be present
in certain string theory models. These strings are electric strings
and are similar to the F-strings discussed in Sec.~\ref{subsec:FD}.

\section{Observation of cosmic strings}
\label{sec:observation}

Previous sections concerned the conditions for cosmic superstrings to be produced and survive long enough to have observable effects. In this section we review the main observational signatures of cosmic strings along with the latest bounds. To minimize the overlap with existing literature, we focus on the more recent work, and particularly the aspects that relate to the string theory origin of cosmic strings.

Cosmic strings are not directly observable, instead their presence can be deduced from their gravitational effects. For example, cosmic string loops decay into gravitational waves~\cite{Vilenkin:1981bx,Hogan:1984is,Vachaspati:1984gt,Bennett:1987vf}, with intense bursts associated with cusps that on average form 
once per oscillation of a loop~\cite{Berezinsky:2000vn,Damour:2000wa}. An entirely different class of signatures of string gravity stems from the peculiar form of their metric. The spacetime around a straight cosmic string is locally flat, but globally conical, with a deficit angle determined by the string tension. This leads to interesting effects, such as wakes of matter forming behind moving strings, line discontinuities in the CMB temperature and polarization, 
and characteristic patterns of lensed images of background light sources. In addition, large matter overdensities created in the wakes at high redshifts can cause early star formation and significantly alter the reionization history of the universe~\cite{rees86,Avelino:2003nn,Pogosian:2004mi}. While such gravitational effects are common to all cosmic strings, they 
can have other signatures in models in which they couple to other forces. 
One example is the electromagnetic radiation from superconducting strings 
studied in \cite{Vilenkin:1986zz,Berezinsky:2001cp,Vachaspati:2008su}. 

Cosmic strings were once thought to be initial seeds for the growth of 
large scale structures in the 
universe \cite{Zeldovich:1980gh,Vilenkin:1981zs}. They could also 
explain the scale-invariant spectrum of large scale CMB temperature 
fluctuations measured by COBE~\cite{cobe}. However, the strings sourced CMB spectrum was in clear disagreement with the sharp peak on the $1^\circ$ scale measured by Boomerang~\cite{boomerang} and Maxima~\cite{maxima} in 2000, which effectively ruled out strings as a significant source of cosmological fluctuations. Subsequently, multiple acoustic peaks in the CMB spectrum detected by WMAP~\cite{wmap} provided convincing evidence of standing waves in the primordial plasma prior to last scattering, which cannot be produced by incoherent active sources, such as strings. Currently, the contribution of strings to the CMB spectrum is limited to be no more than a few percent of the total anisotropy power~\cite{Wyman:2005tu,Bevis:2007gh,Battye:2010hg,Battye:2010xz}. Still, even without existing observational evidence for cosmic strings, the theoretical considerations of previous sections serve as motivation to search for them in the upcoming data. 

Observational constraints on strings are often quoted as a bound on their dimensionless mass per unit length, $G\mu$, which is also the string tension. Such bounds typically assume the scaling configuration in the Abelian Higgs model, where at any time there is roughly one Hubble length string per Hubble volume. 
More generally, the bound on strings depends on the combination of 
$G\mu$ and the string number density $N_s$. In a scaling network, $N_s \sim L^{-2}$, where $L$ is the average interstring distance that remains a fixed fraction of the horizon: $L \sim \xi t$. In the Abelian Higgs model, $\xi \sim {\cal O}(1)$, but can be much smaller in models with lower intercommuting probabilities. Moreover, different types of observations probe different combinations of $\xi$ and $\mu$. As shown in \cite{PogosianTye}, CMB power spectra (and other two-point correlation functions) constrain $\sqrt{N_s}\mu \sim \mu/\xi$, while GW probes essentially constrain the string energy density given by $\mu/\xi^2$. This means that a combination of different probes can, in principle, help to break the degeneracy between a sparse network of heavy strings and a dense network of light strings. As emphasized in \cite{Pourtsidou:2010gu}, this can be particularly relevant for constraining the value of the fundamental string coupling $g_s$, since  in the case of superstring networks, the scaling string configuration is dominated by light and populous F strings at large values of $g_s$, but rare and heavy D and FD strings at small $g_s$.
  
In what follows we will review the latest observational bounds on strings 
coming from different types of observations, with the emphasis on the 
sensitivity of these bounds on the model-dependent details of cosmic 
string networks, especially in connection to the string theoretic origin 
of the cosmic strings. At present, the strongest bounds on the string 
tension come from constraints on the stochastic GW background from pulsar 
timing measurements \cite{Jenet} and LIGO~\cite{Abbott:2009ws}. However, 
these bounds are sensitive to the details of the loop size 
distribution~\cite{Damour:2001bk,Damour:2004kw,Siemens:2006vk,Siemens:2006yp,Olmez:2010bi}. 
Also, the higher dimensional nature of superstrings may imply a lower 
intensity of gravitational waves emitted by 
cusps~\cite{O'Callaghan:2010ww,O'Callaghan:2010sy,O'Callaghan:2010hq}. 
On the other hand, the bounds based on the effects of long strings, such 
as those from CMB, are weaker but also less model-dependent. Importantly, 
the upcoming CMB measurements by Planck and several balloon and ground based 
experiments will produce bounds comparable to those from GW based probes. 
We will also briefly comment on some of the recent ideas for looking for 
cosmic strings with 21 cm and weak lensing surveys.

\subsection{CMB anisotropies}

Cosmic strings induce CMB anisotropies in a way that is fundamentally 
different from the inflationary perturbations. In the latter case, 
inflation sets the initial conditions for the perturbations which then 
evolve forward in time without the production of any additional disturbances.
In contrast, cosmic string networks persist throughout the history of the 
universe and actively source scalar, vector and tensor metric perturbations 
at all times. In particular this means that vector modes -- which quickly 
decay in the absence of a source term and are, for this reason, rarely 
considered in the literature -- are significant for cosmic strings and 
are typically comparable in magnitude to scalar modes. The string generated 
tensor modes are also at a comparable level but their observational impact is generally lower because of the oscillatory nature of gravity waves \cite{Turok:1997gj}. Prior to recombination, density and velocity perturbations of baryon-photon fluid are produced in the wakes of moving cosmic strings, which then remain imprinted on the surface of last scattering.  After recombination, strings crossing our line of sight generate line-like discontinuities in the CMB temperature, which is the so-called Kaiser-Stebbins-Gott (KSG) effect~\cite{Kaiser:1984iv,Gott:1984ef}. Both, wakes and the KSG effect, are induced by the deficit angle in the metric around a string. In addition, matter particles experience gravitational attraction to the string if it is not perfectly straight. The search for cosmic string signatures in the CMB can be broadly divided into attempts to directly detect line discontinuities in the temperature or polarization patterns, and statistical methods based on calculations of various correlation functions.

\subsubsection{Direct searches}

A string passing across our line of sight at any point after 
last scattering would produce a discrete step in the CMB temperature proportional to $G \mu |{\vec v} \times {\hat n}| $, where ${\hat n}$ is the direction of the line of sight. Several groups have tried searching for such line-like features in the existing CMB maps and to forecast the prospects for future observations \cite{Jeong:2004ut,Lo:2005xt,Jeong:2006pi,Jeong:2010ft}. Lo and Wright \cite{Lo:2005xt} employed a digital filter designed to search for individual cosmic strings in the WMAP 1 year data and reported a bound of $G\mu < 1.07 \times 10^{-5}$ which assumes a string moving with velocity $v=1/\sqrt{2}$. They also forecast that Planck will improve on this bound by a factor of two. Jeong and Smoot \cite{Jeong:2006pi} searched for discrete temperature steps in the WMAP 3 year data and arrived at significantly stronger upper limit of $G\mu < 3.7 \times 10^{-6}$ at the 95\% confidence level (CL). Among reasons for this bound being stronger than the one in \cite{Lo:2005xt} is the fact that \cite{Jeong:2006pi} assumes that the number density of strings is approximately known, while the method in \cite{Lo:2005xt} is independent of the string density.  The algorithm of \cite{Jeong:2006pi} was used in \cite{Jeong:2010ft} to study the prospects for direct cosmic string detection with the Planck survey, forecasting a bound of $G\mu<1.5 \times 10^{-6}$ at 95\% CL. The primary limitation on further improving these bounds comes not so much from the instrumental noise and angular resolution, but from the fact that the CMB is dominated by the Gaussian fluctuations on scales comparable to the size of the horizon at decoupling. Amsel and collaborators \cite{Amsel:2007ki} used the Canny algorithm, originally developed as a pattern recognition technique based on detecting edges in images \cite{canny}, to trace the lines in the CMB maps across which the intensity contrast is largest. The method was used in \cite{Stewart:2008zq,Danos:2008fq} to forecast constraints from future CMB experiments that will provide few arcminute resolution maps, such as ACT and SPT. They claim that such experiments will achieve bounds of $G\mu < 3 \times 10^{-8}$ if the Canny algorithm fails to detect sharp edges. In~\cite{Danos:2010gx} it was suggested that detectable sharp edges can also be present in the CMB polarization maps.  As the authors of \cite{Stewart:2008zq,Danos:2008fq,Danos:2010gx} admit, their forecasted bounds are of preliminary nature as they assume idealized line discontinuities produced by straight string segments. Actual strings are not straight, and contain both infinite strings and string loops. It will be interesting to see how well the Canny algorithm can detect strings in more realistic maps, such as the KSG map produced by Fraisse et al~\cite{Fraisse:2007nu}, or the maps of Landriau and Shellard~\cite{Landriau:2010cb}. 

One could go beyond simply searching for line like features in the CMB, and actually try to probe the superstring nature of cosmic strings, such as the presence of Y-junctions. The CMB distortions due to Y-junctions, as well as their lensing effect, were studied in \cite{Brandenberger:2007ae}, and the Canny algorithm was used in \cite{Danos:2009vv} to forecast the ability of future CMB measurements to detect their presence. While the results should be taken as preliminary because of a number of simplifying assumptions, it appears that a direct detection of Y-junctions in the CMB is unlikely when the Gaussian noise from the dominant inflationary contribution is taken into account.

\subsubsection{Angular spectra of temperature anisotropies}

Calculating the spectrum of CMB anisotropies sourced by strings is highly non-trivial. Ideally, one needs to start with an initial configuration of fields and evolve them forward in time in an expanding universe together with all the relevant radiation and matter content in the universe. Because the small size of the core of the strings remains constant, while the universe expands, one quickly runs out of numerical resources even if only trying to track the evolution of one long string. Even in the Nambu-Goto approximation, predicting CMB anisotropies on a sizable patch of the sky is extremely challenging, as it involves tracking the evolution of strings from just before the recombination until today over a large range of scales.

The most complete simulation of CMB maps from cosmic strings in the Nambu-Goto (NG) approximation is by Landriau and Shellard \cite{Landriau:2003xf,Landriau:2010cb}, who include all the relevant CMB physics, including string signatures generated at last scattering. Because it is challenging to achieve high angular resolution on the entire sky, in \cite{Landriau:2010cb} they separately generate a low resolution full sky map, a medium resolution $18^\circ$ map, and a high resolution $3^\circ$ map. They clearly demonstrate the importance of the last scattering surface effects on scales of $400 < \ell < 2000$, as well as the need to properly evolve the network through radiation matter equality. Fraisse et al \cite{Fraisse:2007nu} also employed the Nambu-Goto approximation, focusing on predictions for a $7.2^\circ$ fields of view at an arcminute resolution scale, allowing them to produce spectra up to $\ell =10^4$. They ignore the perturbations induced by cosmic strings on the surface of last scattering, making the simulation significantly less challenging numerically than the one in \cite{Landriau:2010cb}. Their assumption of neglecting the last scattering contribution is valid for $\ell \gtrsim 3000$, but, as the work in \cite{Landriau:2010cb,Pogosian:2008am,Bevis:2010gj} shows, it cannot be ignored for smaller $\ell$'s. The same comment applies to the analytical work in \cite{Yamauchi:2010ms} where the last scattering effects were also neglected.

A principally different approach that did not use the NG approximation was taken by Bevis et al \cite{Bevis:2006mj} who actually evolved the cosmic string field configurations in the Abelian-Higgs model (AH). As this is a numerically challenging endeavour, a number of simplifying approximations were needed to make the calculation of CMB spectra feasible. The fields were separately evolved over limited time ranges in the radiation and matter eras to find the two respective sets of eigenmodes of the unequal time correlators (UETCs) of the string stress energy tensor. Then an interpolation scheme was used to connect the eigenmodes of the two scaling regimes. The scaling of UETCs was then assumed to extend their range to later times. To circumvent the problem of resolving the fixed width core of strings in an expanding background, the core size was allowed to grow with the expansion in a prescribed fashion. Note that this approach does not produce CMB maps, as it is designed to directly calculate the spectra.

Another method, also designed to predict the spectra, but not the map, uses the so-called unconnected segment model (USM) implemented in the publicly available code CMBACT \cite{Pogosian:1999np,cmbact}. In the USM, the string network is represented as a collection of uncorrelated straight string segments, an approximation proposed in \cite{Vincent:1996qr} and adapted for calculation of CMB spectra in \cite{Albrecht:1997nt,ABR99,Battye:1997hu,Pogosian:1999np}. The segments of strings are produced at some early time and given random independent orientations and velocities. At later times, a certain fraction of the number of segments decays in such a way as to match the number density given by a separately provided scaling model. The initial  positions and orientations of the segments are drawn from uniform distributions, and the direction of the velocity is taken to be uniformly distributed in the plane perpendicular to the string orientation (longitudinal velocities are neglected). USM is not a means for gaining new insight into the evolution of cosmic string networks. Instead, it is a tool for evaluating CMB spectra for {\it given} scaling parameters, such as the correlation length and rms velocity. In \cite{Battye:2010xz}, it was shown that the CMB spectra obtained from field theoretical simulations of AH strings \cite{Bevis:2006mj} are reproduced by CMBACT when the one-scale parameters measured in the simulation are used as input. CMBACT is also consistent with the spectrum of \cite{Landriau:2010cb}, although the latter has large statistical error bars.

There is a broad agreement on the general shape of the CMB spectrum induced by local cosmic strings (as opposed to global, which are significantly different), such as a single broad peak at $\ell \sim 300-500$, and a $1/\ell$ fall off for $\ell > 3000$, although all currently used methods of calculating CMB spectra still have serious limitations. While there are some differences in the predicted spectrum shapes, the bound on the allowed fraction of the string contribution to the CMB temperature anisotropy is quite consistent between the different groups and is currently at about $10$\% \cite{Wyman:2005tu,Bevis:2007gh,Battye:2010hg,Battye:2010xz}. Landriau and Shellard \cite{Landriau:2010cb} do not fit their spectrum to the WMAP data, but comment that they expect their bound to be similar to that in \cite{Battye:2010hg} where CMBACT was used.

While different groups agree remarkably well on the allowed fraction of string sourced anisotropy, the corresponding bounds on $G\mu$ vary by factors of $2$ or more. This can be attributed to the different effective string densities in the different models, which depend on the details of the modelling.   For conventional strings, with roughly a few Hubble size long strings per Hubble volume at any time, the $10$\% bound on the string induced fraction of the CMB anisotropy translates into $G\mu \lesssim 6 \pm 3 \times 10^{-7}$ when averaged over the models considered in the literature. Planck will significantly reduce the limit on the allowed string contribution and should produce a bound around $0.1$\% based on the temperature anisotropy and the E-mode polarization data \cite{Battye:2007si}.

\subsubsection{B-mode polarization}

Although it has been established that a network of cosmic strings cannot source the majority of the temperature anisotropy \cite{Albrecht:1997nt}, the CMB can still provide a distinctive signature of their presence through the specific primordial B-mode polarization spectrum \cite{Seljak:1997ii,Hu:1997hp,Battye:1998js,PogosianTye,SelSlo,Bevis:2007qz,Pogosian:2007gi,Urrestilla:2008jv,Mukherjee:2010ve}. While intensity gradients automatically generate parity-even, or $E$-mode, patterns in CMB polarization maps, parity-odd, or $B$-mode, patterns are not produced unless there are metric perturbations that can locally have a non-vanishing handedness. Local departures from zero handedness can be due to tensor modes, or gravity waves, which can be represented as linear combinations of left- and right-handed polarizations, as well as due to non-zero vector modes, or vorticity. The B-mode from strings is primarily generated by vector modes, with a spectrum that is different from the one generically produced from tensor modes arising in inflationary scenarios. Future CMB polarization experiments should be able to reveal the presence of cosmic strings through their B-mode signature even if strings contribute as little as $0.1\%$ to the CMB temperature anisotropy \cite{SelSlo,Bevis:2007qz,Pogosian:2007gi,Urrestilla:2008jv,Mukherjee:2010ve}.
 
The string induced B-mode spectrum has two peaks: a small one at $\ell \sim 10$ and a prominent one at $\ell \sim 600-1000$. The less prominent peak at lower $\ell$ is due to rescattering of photons during reionization, which is thought to have happened in the redshift range of $7 < z < 12$. The main peak, at higher $\ell$, is the contribution from the last scattering surface. Both peaks are quite broad because a string network seeds fluctuations over a wide range of scales at any given time.
The position of the main peak is determined by the most dominant Fourier mode stimulated at last scattering. One can estimate this dominant scale using simple analytical considerations based on the uncorrelated segment picture presented in the previous section. It primarily depends on the string correlation length and the average string velocity at last scattering. Measuring the location of the main peak would provide valuable information about the properties of the cosmic string network, and give us a clue about their origin.

\begin{figure}
\begin{center}
  \includegraphics[width=3.5in]{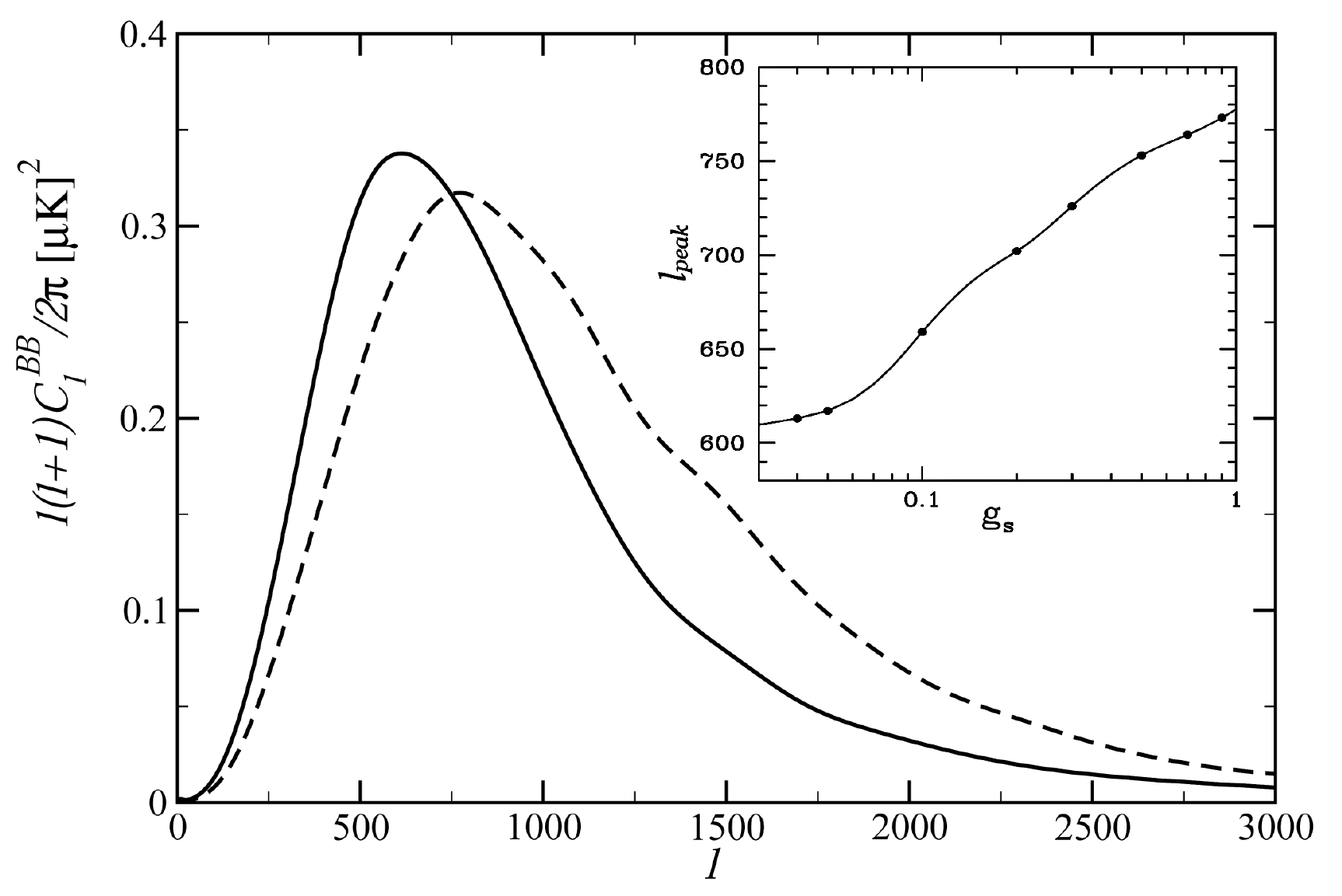}
\end{center}
\caption{The B-mode spectra for two values of $g_s$ from \cite{Pourtsidou:2010gu}. The insert shows a plot of the peak position vs $g_s$.
}
\label{fig:bmode}
\end{figure}

Pourtsidou et al \cite{Pourtsidou:2010gu} have studied the B-mode spectra sourced by
multi-tension cosmic superstring networks. They discovered that, depending on the magnitude of the fundamental string coupling $g_s$, the evolution of F-D networks falls into one of two distinct scaling regimes. In one of them, corresponding to $g_s$ of order unity, the networkÕs power spectrum is dominated by populous light F and D strings, while in the other regime, at smaller values of $g_s$,  the spectrum is dominated by rare heavy D strings. The two regimes result in different locations of the main peak in the B-mode polarization spectrum, as can be seen in Fig.~\ref{fig:bmode}. Thus, measuring the peak would tell us something about the value of $g_s$.  The analysis in \cite{adam-in-prep} shows that Planck~\cite{planck} and SPIDER~\cite{spider} will not be able to distinguish between the two regimes even if the string fraction is close to the current WMAP limit. However, the advertised sensitivities of QUIET~\cite{quiet} and PolarBear~\cite{polarbear} experiments could allow them to distinguish between the two regimes if strings turn out to contribute more than 1\% to the CMB temperature anisotropy. The proposed fourth generation CMB satellites, such as CMBPol~\cite{Baumann:2008aq} and COrE \cite{Collaboration:2011ck} would improve this by roughly an order of magnitude and hence will be a powerful observatory for testing fundamental physics. 

Mukherjee et al \cite{Mukherjee:2010ve} recently analyzed the prospects of detecting and distinguishing topological defects in future data from CMBPol. In particular, they considered spectra produced by inflationary gravitational waves, textures and the Abelian-Higgs cosmic strings of \cite{Bevis:2006mj}. They found that cosmic strings can be detected and correctly identified at $3\sigma$ level (as different from GW or textures) if their contribution to the CMB temperature spectrum is only $0.2$\%.

\subsubsection{Non-gaussianity}

Cosmic strings are extended objects evolving non-linearly under the force of their own tension and generating metric and density perturbations that are intrinsically non-Gaussian. The central limit theorem can make detecting their non-Gaussian signal challenging on scales that were affected by many strings. Experimental noise and the dominant Gaussian inflationary perturbations can further complicate the search. In principle, there are many ways in which the non-Gaussianity of a random process can be manifested. Having one test (e.~g. the bispectrum) come out negative, does not imply that another test (e.~g. the trispectrum) would not turn out to be positive. Several groups have made predictions for various non-Gaussian estimators that could be sensitive to cosmic strings but, 
so far, there has been no significant detection of cosmological non-Gaussianity of any kind.

Using an analytical model of the string network, Moesner et al.~\cite{Moessner:1993za} studied the kurtosis of CMB temperature gradient maps. They found the difference between the stringy and inflationary value for the kurtosis to be inversely proportional to the angular resolution and to the number of strings per Hubble volume. In \cite{Gilbert:1995de}, Gilbert and Perivolaropoulos used Monte Carlo simulations of perturbations induced by cosmic strings to show that the non-Gaussian signatures of the string patterns are detectable by tests based on the moments of the distributions only for angular scales smaller than a few arcminutes and for maps based on the gradient of temperature fluctuations. Avelino et al~\cite{Avelino:1998vu} used high-resolution numerical simulations to compute the one-point probability density function of the matter density field, as well as its skewness, kurtosis, and genus curves for different smoothing scales. They concluded that on scales smaller than $\sim 1$Mpc, perturbations seeded by cosmic strings are very non-Gaussian. Gangui et al~\cite{Gangui:2001fr} used the USM to calculate the angular bispectrum of CMB anisotropies for the equilateral configuration of multipoles and concluded that large statistical errors make the signal unobservable for $\ell<1000$. However, that work neglected the vector mode contribution, plus the signal could be larger for other configurations and on smaller scales.

More recently, the CMB bispectrum induced by strings on small angular scales was studied analytically by Hindmarsh et al~\cite{Hindmarsh:2009qk} under the assumption that anisotropies are primarily due to the KSG effect. They found that isosceles configurations lead to a negative bispectrum with a power law decay $\ell^{-6}$ for large $\ell$, while collapsed triangles are associated with a positive bispectrum, and squeezed triangles also exhibit negative values. In \cite{Hindmarsh:2009es} similar methods were used to calculate the CMB trispectrum from strings. It was found that the trispectrum is predicted to decay like a power-law $\ell^{-\rho}$ with exponent $6<\rho<7$ depending on the string microstructure. They explored two classes of wavenumber configuration in Fourier space, the kite and trapezium quadrilaterals, and found that the trispectrum can be of either sign and is enhanced for squeezed quadrilaterals. The work in \cite{Hindmarsh:2009qk,Hindmarsh:2009es} did not study the detectability prospects.

Regan and Shellard~\cite{Regan:2009hv} also used analytic calculations of the KSG effect to estimate the CMB power spectrum, bispectrum and trispectrum. Their analysis focused on predictions for the  forthcoming CMB experiments, and specifically for the Planck satellite. They found a particular shape for the string bispectrum which is clearly distinguishable from inflationary bispectra. They estimate that the nonlinearity parameter $f_{NL}$ often used to characterize the bispectrum is $-20$ for a string contribution that is consistent with current CMB data. They also calculate the trispectrum for parallelogram configurations on angular scales relevant for WMAP and Planck, as well as on very small angular scales. Interestingly, they find that, while the bispectrum is suppressed by symmetry considerations, the cosmic string trispectrum is large. In their estimate, the trispectrum parameter is $\tau_{NL}\sim 104$, and can provide strong constraints on cosmic strings as observational estimates for the trispectrum improve.

\subsection{Gravitational waves}

As Damour and Vilenkin have shown in \cite{Damour:2001bk}, the 
stochastic gravitational wave (GW) background generated by oscillating 
loops of cosmic strings is strongly non-Gaussian, and includes occasional 
sharp bursts due to cusps and kinks that stand above the ``confusion'' 
GW noise made of many smaller overlapping bursts. They have argued that 
even if only 10\% of all string loops have cusps,  then LISA would detect 
string tensions as small as $G \mu \sim 10^{-13}$. They have also shown 
that the constraints on $G \mu$ from pulsar timing experiments become 
much stronger when the effect of cusps and kinks is taken into account.  
In a follow up paper~\cite{Damour:2004kw}, they have considered GW bounds 
on networks of cosmic superstrings allowing for smaller reconnection 
probabilities, as well as allowing for the length of newly formed loops 
to be a free parameter.

Siemens et al~\cite{Siemens:2006vk,Siemens:2006yp,Olmez:2010bi} generalized the derivation in \cite{Damour:2001bk,Damour:2004kw} to include the effects of late time acceleration of the expansion, and to allow for arbitrary cosmic string loop distributions and found somewhat lower burst rates than previous estimates. They went on to analyze constraints on strings from current and planned gravitational wave detectors, as well as from big bang nucleosynthesis (BBN), CMB, and pulsar timing constraints. In the case when loops are large at formation (the loop size parameter $\alpha=0.1$), they find that pulsars \cite{Jenet} currently provide the tightest bounds of $G\mu \lesssim 10^{-9}$ for $p=1$, and $G\mu < 10^{-12}$ for $p < 10^{-2}$, where $p$ is the intercommutation probability \cite{Olmez:2010bi}. Battye and Moss~\cite{Battye:2010xz} also analyzed the pulsar timing bounds on $G\mu$ for various values of $\alpha$, using the data analysis of \cite{Jenet}. They report a bound of $G\mu< 7\times 10^{-7}$ in the limit $\alpha \ll 60G\mu$, and $G\mu< 5 \times 10^{-11}/\alpha$ when $\alpha \gg 60G\mu$. The large $\alpha$ limit is favoured by the most recent (and the largest) simulation of loop formation by Blanco-Pillado et al \cite{BlancoPillado:2011dq} which suggests $\alpha \sim 0.1$.

Recently, O'Callaghan et al \cite{O'Callaghan:2010ww,O'Callaghan:2010sy} 
explored the kinematical effect of extra dimensions on the gravity wave 
emission from cosmic string cusps. They discovered that additional 
dimensions reduce the probability of cusp formation, as well as lead 
to smoothing of the cusps. This results in a significant damping on 
the gravity waves emitted by cusps, and thus significantly relaxes pulsar 
timing bounds on cosmic superstrings. As discussed in 
\cite{BlancoPillado:2000xy,BlancoPillado:2005jn}, the extra dimensions 
can also be viewed as additional degrees of freedom living on the 
strings, meaning that there are currents that can round off the cusps. 
O'Callaghan and Gregory also considered the effects of extra dimensions 
on GW from kinks \cite{O'Callaghan:2010hq}. They found that while the 
signal is suppressed, the effect is less significant than that for cusps. 
On the other hand, Binetruy et al~\cite{Binetruy:2010cc} find that 
junctions on cosmic superstring loops give rise to the proliferation of 
sharp kinks.

\subsection{Lensing of compact light sources}

The peculiar form of the metric around a cosmic strings can result 
in characteristic lensing patterns of distant light 
sources~\cite{Hogan_84,Gott:1984ef,Vilenkin_86}. For instance, a straight 
long string passing across our  line of sight to a distant galaxy can 
make it look as two exact copies of the same galaxy. In a more general 
case of loops and non-straight strings, the patterns will be more 
complicated, but still have a characteristic stringy 
signature~\cite{deLaix:1996vc,deLaix:1997jt,Sazhin:2006kf}.

When strings bind and create junctions, as in the case of F-D superstring networks, the resulting configurations can lead to novel gravitational lensing patterns. Shlaer and Wyman~\cite{Shlaer:2005ry} used exact solutions to characterize these phenomena, one example being the tripling of images when lensed by a Y-junction. In a related work, Brandenberger et al~\cite{Brandenberger:2007ae} derived the metric of a static string junction, with an arbitrary number of strings joining. They have shown that the metric is flat away from the strings, yet each string segment produces a deficit angle, and thus can deflect both light and matter. They also find that junctions lead to a characteristic pattern of multiple lensed images, and that a uniformly moving string junction produces a junction of line discontinuities in the CMB tempertature.

Mack et al~\cite{Mack:2007ae} argued that the existence of cosmic strings 
can be strongly constrained by the next generation of gravitational lensing 
surveys at radio frequencies.  Using simple models of the loop population 
they find that existing radio surveys such as CLASS have already ruled out 
a portion of the cosmic string model parameter space, while future 
interferometers, such as LOFAR and SKA, can give an upper bound of 
$G\mu/c^2 < 10^{-9}$, which is tighter than current constraints from pulsar 
timing and the CMB by up to two orders of magnitude. Somewhat less optimistic 
conclusions were reached by Kuijken et al~\cite{Kuijken:2007ma}, who also 
studied the gravitational microlensing of distant quasars by cosmic 
strings. They noted that such events will have a characteristic light 
curve in which a source would appear to brighten by exactly a factor of 
two before reverting to its original apparent brightness. They find that 
with limits on the density of cosmic strings from the CMB fluctuation 
spectrum one is left with only a small region of parameter space (in 
which the sky contains about $3\times10^5$ strings with deficit angle of 
order 0.3 milli-arcseconds) for which a microlensing survey of exposure 
$10^7$ source-years, spanning a 20--40-year period, might reveal the 
presence of cosmic strings. 

The effect of loop clustering on microlensing, as well as the effect 
of gravitational lensing due to a moving string on pulsar timing was 
studied in \cite{Pshirkov:2009vb}. Quasar variability has also been 
used to constrain the cosmic string density \cite{Tuntsov:2010fu}.

\subsection{Cosmic rays}
\label{subsec:crays}

Cosmic strings typically move at relativistic speeds and the root-mean-square
speed in the center of mass frame of a loop of string is $1/\sqrt{2}$.
Occasionally there are points on the string, called ``cusps'', that can
become ultra-relativistic. Also, there can be sharp features on strings,
called ``kinks'', that propagate along the string with phase velocity
equal to the speed of light. These distinctive features are of interest
because they can emit beams of a variety of forms of radiation which
can potentially be detected on Earth as cosmic rays.

Several authors have calculated the emission of particles from strings
and the possibility of detecting them as cosmic rays (for a review and
early references see \cite{Bhattacharjee:1998qc}).
An important feature is that the flux of particles on Earth is
{\it inversely} related to the string tension. Thus lighter strings
produce larger cosmic ray fluxes. The reason is simply that the density
of string loops is greater if the strings are lighter, and the larger
number of strings give a larger cosmic ray flux. Hence, if there are
cosmic strings that emit cosmic rays, the constraints imply a {\it lower}
bound on the string tension.

Another important constraint on the cosmic string scenario arises
because the particles emitted by strings generally include protons and
also very high energy ($\sim 10^{20}$ eV) photons \cite{Aharonian:1992qf}.
Even though the nature of the ultra-high energy cosmic rays is not clear
at present - they could be protons or heavy nuclei or an admixture - it
is certain that they do not include a significant photon component. Recent
analyses show that strings in models with particular interactions may be
able to source the ultra-high energy cosmic rays without conflicting with
the photon bounds \cite{Vachaspati:2009kq}.
In these models, the string tension is also bounded from below.

Since string theories generically contain a
dilaton and other moduli, these will be radiated
and can provide stringent constraints. The case when the dilaton has
gravitational-strength coupling to matter has been discussed in 
\cite{Damour:1996pv}, with constraints arising from a number of 
different experiments and observations. The constraints are
obtained in the two-dimensional parameter space given by the 
cosmic string tension and the mass of the dilaton. In the case
of large volume and warped Type-IIB compactifications, the coupling
of the moduli is stronger than gravitational-strength, and the
resulting constraints in the three dimensional parameter space -- 
cosmic string tension, moduli mass, coupling strength -- have been 
analyzed in \cite{Sabancilar:2009sq}. String theory cosmic strings 
can also be expected to provide distinctive cosmic ray signatures
via the moduli emitted from cusps. 

Recently, there has also been discussion of the gravitational coupling
between photons and cosmic strings and the consequent emission of light
from strings \cite{Garriga:1989bx,JonesSmith:2009ti,Steer:2010jk}.
This particular signature is generic to cosmic strings,
whether or not they originate from string theory.

\subsection{Other proposals}

While GW, CMB, strong lensing and large scale redshift surveys have long been used to search cosmic strings, the prospects of future 21 cm and large scale weak lensing surveys motivated investigations of their ability to constrain cosmic strings.

Neutral hydrogen absorbs or emits 21 cm radiation throughout all times after recombination. Thus, at least in principle, it can be used to map the distribution of matter in the dark ages. Cosmic strings would stir the hydrogen as they move around and create wakes, leading to 21 cm brightness fluctuations. Khatri and Wandelt~\cite{Khatri:2008zw} calculated the contribution of strings to the cosmic 21 cm power spectrum at redshifts $z > 30$. They find that under certain optimistic assumptions, future experiments with a collecting area of $10^4-10^6$ km$^2$, can in principle constrain cosmic strings with tension $G\mu$ in the $10^{-10}-10^{-12}$ range. The same strings that create wakes would also perturb the CMB via the KSG effect, leading to spatial correlations between the 21cm and CMB anisotropies. Berndsen et al~~\cite{Berndsen:2010xc} calculated the CMB/21 cm cross-correlation due to this effect and evaluated its observability. Brandenberger et al~\cite{Brandenberger:2010hn} noted that the ionized fraction in the cosmic string wake is enhanced, leading to an excess 21 cm radiation confined to a wedge-shaped region. Hernandez et al~\cite{Hernandez:2011ym} estimated the angular 21 cm power spectrum from a scaling network of strings. It remains to be seen if terrestrial and galactic foregrounds (which become very bright at low frequencies) can be overcome to use 21 cm for mapping the high redshift distribution of matter.

Thomas et al~\cite{Thomas:2009bm} pointed out a new interesting effect that vector perturbations sourced by strings or other topological defects can have on weakly lensed images of distant galaxies. Namely, defects can generate a curl-like (or B-mode) component in the weak lensing signal which is not produced by standard density perturbations at linear order. They argue that future large scale weak lensing surveys should be able to detect this signal even for string tensions an order of magnitude lower than current constraints.

\section{Summary}
\label{sec:summary}

We have discussed the existence of cosmic string solutions in string
theory and the variety of ways to observe them. The existence and 
properties of the cosmic string network depend on the details of 
the string theory scenario that is used to describe our particular
patch of the cosmos. If we are lucky and the conditions are right, 
a network of string theory cosmic strings may be present in our
visible universe. Further, the structure of the string network
is similar to a web where three strings can come together to
form a ``Y'' junction.
The web itself is not unique to string theory cosmic strings: 
certain field theory models can also lead to a web {\it e.g.} as
for $Z_3$ strings. However, the suppressed intercommutation
probability of string theory strings has not been seen in any field
theory and this can potentially lead to a unique signature
of string theory, since a lower intercommutation rate leads to
a higher density of strings.

We have described the growing set of observational tools that can
be used to look for cosmic strings. As yet cosmic strings have not
been discovered and their absence can be used to constrain certain
string theory models. The constraints restrict the cosmic string
tension and the energy density in the string network. The limits 
from pulsar timing are
potentially much stronger than those from the CMB, but they are 
also much more model-dependent.
As a rough guide, the string tension $\mu$ is limited to 
$G\mu \lesssim 6 \times 10^{-7}$
due to constraints from CMB and the most favourable (assuming that cosmic string loops are tiny) interpretation 
of the pulsar timing bounds. In the case when there is a significant population of large loops, as favoured by recent numerical simulations, 
the pulsar bound becomes tighter by two orders of magnitude.
These constraints will get significantly stronger with upcoming 
observational efforts.
The hope, of course, is that these efforts will actually detect a 
cosmic string network and lead to observational evidence for string 
theory!

\

{\it Acknowledgments} We thank Tasos Avgoustidis, Richard Battye, 
Jose Blanco-Pillado, Adam Moss, Alkistis Pourtsidou, Ben Shlaer, 
Xavier Siemens, Dani Steer and Alex Vilenkin for useful comments 
and discussions.

\end{document}